\documentclass[aps,preprint]{revtex4}%
\usepackage{amsfonts}
\usepackage{amsmath}
\usepackage{amssymb}
\usepackage{graphicx}%
\begin{document}
\title{Antiferroelectric resonance in noncentrosymmetric multi-sublattice magnets}
\author{V. N. Krivoruchko and D. A. Yablonskii}
\affiliation{Donetsk Physics \& Technology Institute Ukrainian SSR Academy of Sciences,
Str. R. Luxemburg 72, 83114 Donetsk, Ukraine}
\date{\today}


\begin{abstract}
We predict the phenomenon of antiferroelectric resonance (AFER), in which an
\textit{ac} electric field causes magnetic ions located at noncentrosymmetric
positions in a multi-sublattice magnet to undergo magnetic transitions
corresponding to exchange collective excitations of the system. We construct a
theory of such resonances, and show that in magnets with collinear magnetic
structures AFER is caused by relativistic and exchange-relativistic
magnetoelectric interactions, while in the noncollinear magnets a significant
contribution can also come from the exchange magnetoelectric interaction. We
predict an exchange enhancement of the resonance by exchange modes, and
discuss the role of AFER in causing resonant enhancement of magnetooptical
phenomena. The main results of our theory are illustrated with the
four-sublattice rhombohedral antiferromagnets $\alpha$-Fe$_{2}$O$_{3}$ and
Cr$_{2}$O$_{3}$, as examples.

Journal-ref: Zh. Eksp. Teor. Fiz. 94, 268-276 (1988); [Sov. Phys. JETP 67 (9)
1886--1890 (1988)]

\end{abstract}
\maketitle

\section{Introduction}

The traditional method of investigating the resonance properties of
magnetically-ordered crystals consists of subjecting them to a high-frequency
(HF) magnetic field. Using this method we can reliably identify both acoustic
modes (AM) and, in special cases, exchange modes (EM) of the crystal
\cite{1,2,3,4,5,6}. Excitation of EM with an HF magnetic field is possible
only if the former are magnetically-active (i.e., coupled to oscillations of
the ferromagnetism vector \textbf{M} of the system). However, the coupling of
the EM to \textbf{M} comes about through relativistic and
exchange-relativistic interactions and therefore the intensity of absorption
by EM is weakened compared to AM \cite{5,6,7}. For nonmagnetically-active EM
interaction with a magnetic field is impossible, by virtue of selection rules
(for example, in the case of EM which is odd under inversion).

In this paper we show that a natural way to resonantly excite magnetic-system
exchange modes which are odd under inversion is to act on them with a HF
electric field \cite{8}. We construct a theory which describes the excitation
of such modes by an \textit{ac} electric field. The corresponding resonance
will be referred to as antiferroelectric (AFER), because the EM are
oscillations of the antiferromagnetism vectors \textbf{L}$_{i}$. It is
significant that in AFER the HF electric field is directly connected with the
vectors\textbf{ L}$_{i}$, (our discussion centers on transitions directly
induced by the \textit{ac} electric field between the magnetic levels of the system).

The general condition for the existence of AFER will be formulated below. Here
we only remark that in the majority of cases a sufficient condition for the
existence of AFER in noncentrosymmetric crystals is the presence of magnetic
ions at the noncentrosymmetric positions. Examples of such magnets are
hematite, iron garnets, ferrite spinels, orthoferrites, and other compounds.

We will illustrate the essential features of AFER in noncentrosymmetric
crystals with the examples of $\alpha$-Fe$_{2}$O$_{3}$ and Cr$_{2}$O$_{3}$,
where the former (according to Turov \cite{9}) is even under inversion, while
the latter is odd.

\section{Antiferromagnetic resonance in hematite}

As an example of a noncentrosymmetric crystal with a magnetic structure which
is even under inversion, let us discuss the four-sublattice rhombohedral
antiferromagnet $\alpha$-Fe$_{2}$O$_{3}$ (hematite). Following Ref. \cite{10},
we introduce the following linear combinations of the sublattice
magnetizations \textbf{M}$_{i}$, ($i$ = 1, 2, 3, 4):%
\begin{equation}
\mathbf{M}=\mathbf{M}_{1}\mathbf{+M}_{2}\mathbf{+M}_{3}\mathbf{+M}%
_{4}\mathbf{=}4M_{o}\mathbf{m}\text{ , \ }\mathbf{L}_{1}=\mathbf{M}%
_{1}-\mathbf{M}_{2}-\mathbf{M}_{3}\mathbf{+M}_{4}\mathbf{=}4M_{o}%
\mathbf{l}_{1}\text{,\ }%
\end{equation}%
\begin{equation}
\mathbf{L}_{2}=\mathbf{M}_{1}-\mathbf{M}_{2}\mathbf{+M}_{3}-\mathbf{M}%
_{4}\mathbf{=}4M_{o}\mathbf{l}_{2}\text{, \ \ }\mathbf{L}_{3}=\mathbf{M}%
_{1}\mathbf{+M}_{2}-\mathbf{M}_{3}-\mathbf{M}_{4}\mathbf{=}4M_{o}%
\mathbf{l}_{3}\text{,} \tag{1}%
\end{equation}
where \textbf{M}$_{o}$ is the magnitude of the sublattice magnetization. In
Table I we show the classification of the vectors (1) relative to the
irreducible representations of the group $D_{3d}^{6}$, (for character tables
of these irreducible representations and their notation see Ref. \cite{11}).
The vectors \textbf{m} and \textbf{l}$_{1}$, generate even, and \textbf{l}%
$_{2}$, and \textbf{l}$_{3}$, odd, representations of this group. The system
Hamiltonian has the form \cite{10}:%
\[
(4M_{o})^{-1}W=-\frac{1}{2}H_{e1}\mathbf{l}_{1}^{2}+\frac{1}{2}H_{e2}%
\mathbf{l}_{2}^{2}+\frac{1}{2}H_{e3}\mathbf{l}_{3}^{2}+\frac{1}{2}%
H_{eo}\mathbf{m}^{2}+\frac{1}{2}H_{A1}l_{1z}^{2}+\frac{1}{2}H_{A2}l_{2z}^{2}+
\]%
\begin{equation}
+\frac{1}{2}H_{A3}l_{3z}^{2}++\frac{1}{2}H_{Ao}m_{z}^{2}+H_{D}(m_{y}%
l_{1x}-m_{x}l_{1y})+H_{D}^{^{\prime}}(l_{2x}l_{3y}-l_{2y}l_{3x})-(\mathbf{H}%
+\mathbf{h})\mathbf{m}%
\end{equation}
Here $H_{ei}$ is the effective field of the exchange interaction, $H_{Ai}$ is
the anisotropy field, and $H_{D}$ , and $H_{D}^{^{\prime}}$ are the
Dzyaloshinskii fields.

In the approximation linear in the electric field \textbf{E}(t) we must add to
(2) the invariants (see Table I)%
\[
(4M_{o})^{-1}\mathbf{EP}=E_{x}\{R_{3}(l_{1z}l_{3y}-l_{1y}l_{3z})+R_{2}%
(l_{1z}l_{2x}-l_{1x}l_{2z})+r_{3}(m_{x}l_{3z}-m_{z}l_{3x})\}
\]%
\[
+E_{y}\{R_{3}(l_{1x}l_{3z}-l_{1z}l_{3x})+R_{2}(l_{1z}l_{2y}-l_{1y}%
l_{2z})+r_{3}(m_{y}l_{3z}-m_{z}l_{3y})\}
\]%
\begin{equation}
+E_{z}\{\Pi_{1}\mathbf{ml}_{3}+\Pi_{2}\mathbf{l}_{1}\mathbf{l}_{2}%
+R_{3z}(l_{1y}l_{3x}-l_{1x}l_{3y})+r_{2z}(m_{x}l_{2y}-m_{y}l_{2x})\}\text{ .}%
\end{equation}
The antiferroelectric constants $\Pi_{i}$ have their origin in exchange, while
$R_{i}$ and $r_{i}$ are due to relativistic-exchange effects. We neglect terms
of relativistic origin.

The dynamic properties of the system are described by the equations
\cite{1,2}: $i$\textbf{\.{m}} = [\textbf{m}, $W$] and $i\mathbf{\dot{l}}_{\nu
}$ = [\textbf{l}$_{\nu}$ ,$W$], $\nu$ = 1, 2, 3. Linearizing these equations,
we find that the AM correspond to oscillations of the vectors \textbf{m} and
\textbf{l}$_{1}$, which are even under inversion; these oscillations couple to
the external magnetic field, i.e., they are magnetic dipole-active. The EM
correspond to oscillations of the vectors \textbf{l}$_{2}$ and \textbf{l}%
$_{3}$ which do not interact with an external magnetic field or with the AM
because they are odd under inversion. At the same time, an electric field
couples to the vectors \textbf{l}$_{2}$ and \textbf{l}$_{3}$ (see Table I),
i.e., the EM are electric dipole-active. These properties of the AM and EM are
valid both for the easy-axis and easy-plane phases of hematite (see Table II).

Let us begin with an investigation of the EM in the easy-plane phase with a
constant magnetic field \textbf{H}
$\vert$%
$\vert$
x. Solving the linearized equations of motion, we find that a HF electric
field \textbf{E}(t) $\sim$ \textbf{E}$exp$(i$\omega$t) induces oscillations of
the vectors \textbf{l}$_{2}$, \textbf{l}$_{3}$ $\sim$ $exp$(i$\omega$t) with
amplitudes%
\begin{equation}
l_{3z}=-\gamma^{2}R_{3}(H_{e1}+H_{e2})E_{x}(\omega^{2}-\omega_{o1}^{2})^{-1}%
\end{equation}%
\begin{equation}
l_{2z}=\gamma^{2}\{-R_{3}(H_{e1}+H_{e2})E_{y}+i[R_{3z}+(\Pi_{2}-\Pi
_{1})m]\omega E_{z}\}(\omega^{2}-\omega_{o2}^{2})^{-1}%
\end{equation}%
\begin{equation}
l_{3x}=\gamma^{2}\{[R_{3z}+(\Pi_{2}-\Pi_{1})m](H_{e1}+H_{e2})E_{z}+i\omega
R_{2}E_{y}\}(\omega^{2}-\omega_{o2}^{2})^{-1}%
\end{equation}%
\begin{equation}
l_{2y}=-\gamma^{2}R_{2}i\omega mE_{y}(\omega^{2}-\omega_{o2}^{2})^{-1}\text{
.}%
\end{equation}
In this expression we include only those terms of exchange-relativistic origin
which do not contain $m$; $\gamma$ is the gyromagnetic ratio, and the EM
frequency and magnetization are equal to
\begin{equation}
\gamma^{-2}\omega_{o1}^{2}=(H_{e1}+H_{e3})(H_{e1}+H_{e2})-[(H_{eo}%
-H_{e3})m+H_{D}^{^{\prime}}-H-H_{D}]H_{e1}m
\end{equation}%
\begin{equation}
\gamma^{-2}\omega_{o2}^{2}=(H_{e1}+H_{e3})(H_{e1}+H_{e2}+H_{A2})+(H+H_{D}%
+H_{D}^{^{\prime}})[H+H_{D}+(H_{2e}-H_{eo})m]
\end{equation}%
\[
m=m_{x}=(H+H_{D})(H_{e1}+H_{eo})^{-1}\text{ .}%
\]
We remark that the spectrum of magnetic excitation of 4-sublattice hematite
was calculated in Refs. \cite{12,13,14}; the results derived there coincide
with Eqs. (8) and (9) [see also (17) below]. Inelastic neutron scattering
experiments \cite{12} show that EMs in hematite are located in the infrared band.

To linear approximation in the spin deviations, we have from (3) for the
components of the electric polarization vector:%
\[
P_{x}=-4M_{o}(R_{3}-r_{3}m)l_{3z}\text{, \ \ }P_{y}=4M_{o}R_{2}l_{2z}\text{, }%
\]%
\[
P_{z}=4M_{o}[(\Pi_{2}+r_{2z}m)l_{2y}+(\Pi_{1}m-R_{3z})l_{3x}%
\]
Introducing the electric polarization tensor \textbf{P}($\omega$) = $\alpha
$($\omega$)\textbf{E}($\omega$), we find from Eqs. (4)-(7) and the relations
above that the spin part of the HF electric polarization tensor $\alpha
$($\omega$) in the easy-plane phase with the magnetic field \textbf{H}
$\vert$%
$\vert$
x has the following nonzero components:%
\begin{equation}
\alpha_{xx}(\omega)=4M_{o}\gamma^{2}R_{3}^{2}(H_{e1}+H_{e2})(\omega^{2}%
-\omega_{o1}^{2})^{-1}%
\end{equation}%
\begin{equation}
\alpha_{yy}(\omega)=4M_{o}\gamma^{2}R_{2}^{2}(H_{e1}+H_{e3})(\omega^{2}%
-\omega_{o2}^{2})^{-1}%
\end{equation}%
\begin{equation}
\alpha_{yz}(\omega)=i4M_{o}\gamma^{2}R_{2}[R_{3z}+(\Pi_{2}-\Pi_{1}%
)m]\omega(\omega^{2}-\omega_{o2}^{2})^{-1}%
\end{equation}%
\begin{equation}
\alpha_{zz}(\omega)=4M_{o}\gamma^{2}[R_{3z}+(\Pi_{2}-\Pi_{1})m][(H_{e1}%
+H_{e2})(\Pi_{1}m-R_{3z})](\omega^{2}-\omega_{o2}^{2})^{-1}%
\end{equation}
In Eq. (10) we neglect the term $r_{3}m$ compared to $R_{3}$, while in Eq.
(13) we neglect $r_{2z}m$ compared to $\Pi_{2}$.

For the easy-axis phase ($l_{1z}$ = 1, \textbf{H}
$\vert$%
$\vert$
z), we have%

\begin{equation}
P_{x}=4M_{o}(R_{3}l_{3y}+R_{2}l_{2x})\text{, \ \ }P_{y}=4M_{o}R_{2}%
l_{2y}\text{, \ \ }P_{z}=4M_{o}\Pi_{2}l_{2z}\text{\ .}%
\end{equation}
Calculating the oscillation amplitudes of the antiferromagnetic vectors
\textbf{l}$_{2}$(t) and \textbf{l}$_{3}$(t) under the action of \textit{ac
}electric field \textbf{E}(t) $\sim$ \textbf{E}$exp$(i$\omega$t) in this phase
and substituting them into Eqs. (14), we obtain the following nonzero spin
contributions to the electric polarization tensor:%

\begin{equation}
\alpha_{xx}(\omega)=\alpha_{yy}(\omega)=4M_{o}\gamma^{2}\{R_{3}^{2}%
(H_{e1}+H_{e2})+R_{2}^{2}(H_{e2}+H_{e3})\}(\omega_{o1}\omega_{o2}-\omega
^{2})(\omega^{2}-\omega_{o1}^{2})^{-1}(\omega^{2}-\omega_{o2}^{2})^{-1}%
\end{equation}%
\begin{equation}
\alpha_{xy}(\omega)=-\alpha_{yx}(\omega)=i4M_{o}\gamma^{2}\{R_{3}^{2}%
(H_{e1}+H_{e2})+R_{2}^{2}(H_{e2}+H_{e3})\}2\omega H(\omega^{2}-\omega_{o1}%
^{2})^{-1}(\omega^{2}-\omega_{o2}^{2})^{-1}%
\end{equation}
Now, the EM frequency equals%
\begin{equation}
\gamma^{-1}\omega_{o1,2}=\{(H_{e1}+H_{e3}+|HA1|)(H_{e1}+H_{e2}+|HA1|)-H_{D}%
^{^{\prime}2}\}^{1/2}\pm H\text{ .}%
\end{equation}

It is clear from Eqs. (10) - (13), (15) and (16) that in the easy-plane and
easy-axis phases the residues at the EM poles of the HF electric polarization
tensor are enhanced by exchange. The absorption intensity of the electric
field at these frequencies is determined by the magnitude of the
antiferroelectric constant; in what follows we give a numerical estimate of
this constant.

Hematite is a straightforward example of a multi-sublattice system in which
excitation of EM by a magnetic field is impossible by virtue of general
selection rules, while for an \textit{ac} electric field these rules allow
such excitations. Let us now turn to a different system -- Cr$_{2}$O$_{3}$.

\section{Electric-dipole active vibrations in Cr$_{2}$O$_{3}$}

The crystal Cr$_{2}$O$_{3}$ possesses a magnetic structure which is odd under
inversion \cite{10}. In the magnetic class which includes Cr$_{2}$O$_{3}$,
inversion combines with the time-reversal operation \textbf{\={1}}$\cdot R$,
and these results in a linear magnetoelectric effect \cite{16,17,18}. The
ground state is $A_{1u}$, with \textbf{l}$_{3}$
$\vert$%
$\vert$
z. The magnetic properties of Cr$_{2}$O$_{3}$ are described by the potentials
(2) and (3), in which it is necessary to make the replacements $H_{e1}%
\longrightarrow-H_{e1}$ , $H_{e3}\longrightarrow-H_{e3}$ and $H_{A3}%
\longrightarrow-H_{A3}$ .

In the exchange approximation the acoustic type of oscillations in Cr$_{2}%
$O$_{3}$ correspond to transverse oscillations of the vectors \textbf{l}$_{3}%
$, and \textbf{m}, while those of exchange type correspond to transverse
oscillation of the vectors \textbf{l}$_{1}$, and \textbf{l}$_{2}$. Allowance
for the Dzyaloshinskii interaction of $H_{D}$, and $H_{D}^{^{\prime}}$ in (2)
leads to dynamic coupling of the EM and AM. Without pausing for detailed
calculations we will present the final results.

\textit{A. Acoustic modes}. Accurate to terms of order $H_{D}/H_{e}$ and
$(H_{A}/H_{e})^{1/2}$ inclusively, the frequencies of the AM are equal to%
\begin{equation}
\gamma^{-1}\omega_{A1,2}=\varepsilon_{A}\pm H\text{ ,}%
\end{equation}%
\[
\varepsilon_{A}^{2}=(H_{e3}+H_{eo})\{H_{A3}-H_{D}^{^{\prime}2}(H_{e2}%
+H_{e3})^{-1}\}\text{ .}%
\]
The nonzero transverse components of the HF magnetic susceptibility tensor
have in this frequency interval the form%
\begin{equation}
\chi_{xx}(\omega)=\chi_{yy}(\omega)=4M_{o}\gamma^{2}\{H_{A3}-H_{D}^{^{\prime
}2}(H_{e2}+H_{e3})^{-1}\}(\omega^{2}-\varepsilon_{A}^{2}+H^{2})(\omega
^{2}-\omega_{A1}^{2})^{-1}(\omega^{2}-\omega_{A2}^{2})^{-1}\text{ ,}%
\end{equation}%
\begin{equation}
\chi_{xy}(\omega)=-\chi_{yx}(\omega)=i4M_{o}\gamma^{2}\{H_{A3}-H_{D}%
^{^{\prime}2}(H_{e2}+H_{e3})^{-1}\}\omega H(\omega^{2}-\omega_{A1}^{2}%
)^{-1}(\omega^{2}-\omega_{A2}^{2})^{-1}\text{ .}%
\end{equation}

To linear order we have from (3) the electric polarization%
\[
P_{x}=4M_{o}(r_{3}m_{x}-R_{3}l_{1y})\text{, \ \ \ }P_{y}=4M_{o}(r_{3}%
m_{y}+R_{3}l_{1x})\text{ .}%
\]
The nonzero spin contribution to the HF electric polarization tensor is
described by the expressions%
\begin{equation}
\alpha_{xx}(\omega)=\alpha_{yy}(\omega)=4M_{o}\gamma^{2}r_{3}^{2}%
\{H_{A3}-H_{D}^{^{\prime}2}(H_{e2}+H_{e3})^{-1}\}(\omega^{2}-\varepsilon
_{A}^{2}+H^{2})(\omega^{2}-\omega_{A1}^{2})^{-1}(\omega^{2}-\omega_{A2}%
^{2})^{-1}\text{ ,}%
\end{equation}%
\begin{equation}
\alpha_{xy}(\omega)=-\alpha_{yx}(\omega)=i4M_{o}\gamma^{2}r_{3}^{2}%
\{H_{A3}-H_{D}^{^{\prime}2}(H_{e2}+H_{e3})^{-1}\}\omega H(\omega^{2}%
-\omega_{A1}^{2})^{-1}(\omega^{2}-\omega_{A2}^{2})^{-1}\text{ .}%
\end{equation}

\textit{B. Exchange modes}. To the same accuracy as above we have for the EM
frequencies%
\begin{equation}
\gamma^{-1}\omega_{o1,2}=\varepsilon_{o}\pm H\text{ ,}%
\end{equation}%
\[
\varepsilon_{o}^{2}=(H_{e1}+H_{e3})(H_{e2}+H_{e3})+(H_{e1}+H_{e2}%
+2H_{e3})H_{A3}+H_{D}^{^{\prime}}(H_{D}^{^{\prime}}+2H_{D})(H_{eo}%
+H_{e3})(H_{e2}+H_{e3})^{-1}\text{ .}%
\]
In this frequency interval,%
\begin{equation}
\chi_{xx}(\omega)=\chi_{yy}(\omega)=-4M_{o}\gamma^{2}H_{D}^{^{\prime}2}%
(H_{e2}+H_{e3})^{-1}\times(\omega^{2}-\varepsilon_{o}^{2}+H^{2})(\omega
^{2}-\omega_{o1}^{2})^{-1}(\omega^{2}-\omega_{o2}^{2})^{-1}\text{ ,}%
\end{equation}%
\begin{equation}
\chi_{xy}(\omega)=-\chi_{yx}(\omega)=i4M_{o}\gamma^{2}H_{D}^{^{\prime}%
2}(H_{e2}+H_{e3})^{-1}\times\omega H(\omega^{2}-\omega_{o1}^{2})^{-1}%
(\omega^{2}-\omega_{o2}^{2})^{-1}\text{ .}%
\end{equation}%
\begin{equation}
\alpha_{xx}(\omega)=\alpha_{yy}(\omega)=4M_{o}\gamma^{2}R_{3}^{2}%
(H_{e2}+H_{e3})\times(\omega^{2}-\varepsilon_{o}^{2}+H^{2})(\omega^{2}%
-\omega_{o1}^{2})^{-1}(\omega^{2}-\omega_{o2}^{2})^{-1}\text{ ,}%
\end{equation}%
\begin{equation}
\alpha_{xy}(\omega)=-\alpha_{yx}(\omega)=-i4M_{o}\gamma^{2}R_{3}^{2}%
(H_{e2}+H_{e3})\times\omega H(\omega^{2}-\omega_{o1}^{2})^{-1}(\omega
^{2}-\omega_{o2}^{2})^{-1}\text{ .}%
\end{equation}

Since the vectors (\textbf{l}$_{1}$, \textbf{m}), which are even under
inversion, and the vectors (\textbf{l}$_{2}$, \textbf{l}$_{3}$), which are odd
under inversion, particles in the EM and AM oscillations, these EM and AM can
be excited both by magnetic and electric fields with \textbf{E}, \textbf{h}
$\bot$ z (see Table III). However, from an experimental point of view the
important thing is the magnitude of the absorption by the EM and AM. We obtain
from (19) - (22) and (24) - (27) the following estimate of the susceptibility
near the resonance frequency:
\[
\alpha(\omega_{o})\sim\gamma M_{o}R^{2}(\omega-\omega_{o})^{-1}\text{,
\ \ \ }\chi(\omega_{o})\sim\gamma M_{o}(H_{A}/H_{e})(\omega-\omega_{o}%
)^{-1}\text{ ,}%
\]%
\[
\alpha(\omega_{A})\sim\gamma M_{o}R^{2}(H_{A}/H_{e})^{1/2}(\omega-\omega
_{A})^{-1}\text{, \ \ \ }\chi(\omega_{A})\sim\gamma M_{o}(H_{A}/H_{e}%
)^{1/2}(\omega-\omega_{A})^{-1}\text{\ ,}%
\]
for EM and AM, respectively. The intensity of the absorption of an electric
field by the EM is $(H_{e}/H_{A})^{1/2}$ times larger than absorption by AM.
For a magnetic field, the situation is reversed: the intensity of the
absorption by AM is $(H_{e}/H_{A})^{1/2}$ times larger than that due to the
EM. It is also clear that for $R^{2}>H_{A}/H_{e}$ , excitation of EM by an
electric field is easier than by a magnetic field.

\section{Physical mechanism of AFER}

The physical mechanism which gives rise to the spin Hamiltonian is the same as
the one discussed previously in constructing a theory of the magnetoelectric
effect and of electric effects in paramagnetic resonance, and a theory of
absorption and scattering of light in magnetically-ordered crystals.

A direct indication of the possibility of experimental observation of AFER is
provided by experiments in which an \textit{ac} electric field induces
transitions between magnetic levels of paramagnetic ions in noncentrosymmetric
sites, i.e., transitions from a state (\textit{l}, \textit{m}) to states
(\textit{l}, \textit{m} + 1 ) where \textit{l} is the orbital and \textit{m}
the magnetic quantum number. Such transitions were observed in Refs.
\cite{19,20} (see also Ref. \cite{22} and citations therein). As we have shown
here, in centrosymmetric crystals which have high concentrations of magnetic
ions and magnetic structures which are even under inversion (e.g., of $\alpha
$-Fe$_{2}$O$_{3}$ type) one consequence of these transitions will be magnetic
excitations of exchange type.

The physics of AFER combines the physical mechanisms of electric-dipole
paramagnetic resonance associated with impurity magnetic ions in
noncentrosymmetric position \cite{21,22,23,24} and of absorption and
scattering of light in systems with a high concentration of magnetic ions
\cite{25,26,38}. In systems with magnetic ions occupying centers of inversion,
the electric dipole activity of the magnetic modes can be due to, e.g., the
additional effect of a constant electric field \cite{27}. We emphasize that in
previously-studied magnets \cite{28,29,30,31,32,33} without centers of
inversion the effects investigated were due to coupling of electric and
magnetic subsystems; in contrast, according to the theory constructed in this
paper, AFER in noncentrosymmetric crystals is caused by direct excitation of
exchange-type magnetic oscillations by an \textit{ac} electric field. We shall
determine the antiferroelectric constants for $\alpha$-Fe$_{2}$O$_{3}$, from
experiments on the shift of the paramagnetic resonance lines for Fe$^{3+}$
ions \cite{22,34}. This latter gives an estimate of $\sim$ 10$^{-2}$ for the
single-ion spin-Hamiltonian constant. In systems with a high concentration of
magnetic ions, contributions to the magnetoelectric effects come also from
ion-ion interactions (in particular from exchange and relativistic- exchange
interactions), which in individual cases increase the value of the constant by
an order of magnitude \cite{35,36}. Therefore in $\alpha$-Fe$_{2}$O$_{3}$
(apparently) $R$ is $\sim$ 10$^{-1}$. The value of $\Pi_{i}$ is also an order
of magnitude larger. It should also be noted that the contribution of
invariants of exchange origin is proportional to the magnetization [see Eqs.
(12) and (3)]; therefore, in the canted phases their contribution can
significantly exceed the contribution from the exchange-relativistic
invariants. For quantitative estimates of the antiferroelectric interaction
constants in Cr$_{2}$O$_{3}$ we make use of the results of experimental and
theoretical studies of the magnetoelectric effect in this compound
\cite{18,35,36}. The latter give for the parameters of the spin Hamiltonian
the value \cite{35,36} $\Pi$ $\approx$ 0.5 ; $R$ $\approx$ 0.05.

Taking into account that $H_{A}/H_{e}$ $\sim$ 10$^{-4}$ in Cr$_{2}$O$_{3}$,
the conditions for observation of EM based on the absorption of an electric
field can be more favorable than those based on absorption of a magnetic field.

\section{Linear resonant magnetoelectric effects}

Let us formulate the general conditions for electric-dipole activity of the
magnetic oscillations.

In the general case the EM and AM are of electric-dipole type if the dynamic
compounds of the antiferromagnetism and ferromagnetism vectors which
correspond to them transform according to the irreducible representations of
the unitary subgroup of the system's magnetic symmetry group (i.e., they
transform just like the compounds of the electric polarization vector
\textbf{P}).

Let us investigate in more detail the general conditions for the existence of
AFER in magnetic structures whose ground state in the exchange approximation
is collinear and satisfies the conditions of the Turov classification \cite{9}
(see also Ref. \cite{39}).

In centrosymmetric crystals it is necessary to distinguish between structures
which are even and odd under inversion.

1. The structures \textbf{\={1}}$(+)$. If a system whose structure is even
under inversion contains magnetic ions which are not located at inversion
centers, then in addition to the basic even antiferromagnetism vector
\textbf{L}$_{o}$ there exists at least one other antiferromagnetism vector
\textbf{l} which is odd under inversion. In the thermodynamic potential of
such a magnet it is possible to have invariants of the form%
\begin{equation}
K_{\alpha\beta\gamma}E_{\alpha}L_{o\beta}l_{\gamma}\text{ , \ \ \ \ }%
\alpha\text{, }\beta\text{, }\gamma=x\text{, }y\text{,\ }z\text{ .}%
\end{equation}
In a static electric field, Eq. (28) leads to the appearance of $l_{\gamma
}\sim E_{\alpha}$ . In analogy with the magnetoelectric effect this phenomenon
can be called antiferroelectric (AFEE).

In an \textit{ac} electric field \textbf{E} \symbol{126}$exp$(i$\omega$t), the
relation (28) causes oscillation of the vector \textbf{l}, and for
$\omega=\omega_{o}$ , excitation of EM ($\omega_{o}$ is the exchange frequency).

2. For structures \textbf{\={1}}$(-)$ odd under inversion, it is necessary to
distinguish between two cases according to parity relative to a translation
\textbf{t}.

Systems with \textbf{\={1}}$(-)$ and \textbf{t}$(+)$ pertain to
antiferromagnets whose thermodynamic potentials include an invariant of the
form%
\begin{equation}
K_{\alpha\beta\gamma}L_{o\alpha}M_{\beta}E_{\gamma}\text{ , }%
\end{equation}
where \textbf{M} is the magnetic moment of the system. If in this case there
also an antiferromagnetic vector \textbf{l} exists which is odd under
inversion, then it is also possible to have an invariant of the form (28).
This invariant gives rise to the presence of AFER and AFEE, while (29) makes
possible excitation of AM by an electric field, i.e., magnetoelectric resonance.

In systems with \textbf{\={1}}$(-)$ and \textbf{t}$(-)$ (the magnetic unit
cell is larger than the crystallographic unit cell) the invariant (29) is
forbidden and there is no static magnetoelectric effect. However, an invariant
of the form (28) is possible, where now the antiferromagnetic vector
\textbf{l} is odd under translation and even under inversion. In this system
there will be both AFEE and AFER.

\section{Conclusion}

We shall dwell in somewhat greater detail on the possibility of experimentally
observing antiferroelectric resonance with EM and AM.

Apparently, low-dimensionality magnets are most convenient from an
experimental point of view for observing EM by resonant methods. In such
systems, because of the presence of weak exchange interactions, the EM and AM
frequencies are comparable as a rule \cite{3, 6}. Examples of
low-dimensionality magnet in which it is possible to observe AFER are the
eight-sublattice antiferromagnets CsMnCl$_{2}\cdot$2H2O and CuCl$_{2}\cdot
$C$_{4}$H$_{8}$SO. We note that EMs were recently observed in the AFMR
spectrum of CuCl$_{2}\cdot$C$_{4}$H$_{8}$SO (see Ref. \cite{40}). As our
analysis shows, in this compound the EM can also be observed in the AFER
spectrum by placing the sample at an electric field antinode within a cavity.
The AM can be observed by using the same method.

The EM frequencies of $\alpha$-Fe$_{2}$O$_{3}$ and Cr$_{2}$O$_{3}$ are located
in the infrared wavelength band. Therefore the features of experimental
observation of AFER in these compounds are close to those in optical
experiments \cite{26}. If the dimensions of the sample are comparable with or
smaller than the wavelength of the \textit{ac} electric field, then the theory
developed above is directly applicable.

For bulk samples of $\alpha$-Fe$_{2}$O$_{3}$ and Cr$_{2}$O$_{3}$, the
equations of motion for the vectors (1) must be considered jointly with the
Maxwell equations. In this case, the exchange spin modes and the
electromagnetic oscillations are found to be coupled, which leads to their
mutual restructuring. A detailed analysis of this question is outside the
framework of the present communication; therefore we will only pause briefly
to treat a specific case.

For \textbf{E} $\perp$ \textbf{k}
$\vert$%
$\vert$
z the dispersion relation or coupled right-hand-polarized electromagnetic
waves and EM $\omega_{o1}$ (17) in hematite has the form%
\begin{equation}
\left[  \left(  \frac{kc}{n_{+}}\right)  ^{2}-\omega^{2}\right]  \left(
\omega_{o1}-\omega\right)  \left(  \omega_{o2}-\omega\right)  =\frac{4\pi
}{n_{+}^{2}}\alpha^{\prime}\omega^{2}\text{ ,}%
\end{equation}
Here $c$ is the velocity of light while the parameter $\alpha^{\prime}$ which
determines the value of the coupling of the branches is%
\[
\alpha^{\prime}=4M_{o}\gamma^{2}\{R_{3}^{2}(H_{e1}\mathbf{+}H_{e2})+R_{2}%
^{2}(H_{e2}+H_{e3})\}\text{ ,}%
\]
$n_{+}$ is the refractive index of the medium without allowance for the EM
contribution. The dispersion of the EM in this frequency band is not
significant and we disregard it. For left-hand-polarized waves the dispersion
relation is obtained from (30) by the replacements $\omega_{o1}$
$\longleftrightarrow$ $\omega_{o2}$ , $n_{+}$ $\longleftrightarrow$ $n_{-}$ .

As is clear from (19)-(22) and (24)-(27), when magnetic oscillations can be
excited by an electromagnetic field it is possible to have resonance
singularity both in the components of the magnetic susceptibility tensor and
in the components of the electric polarization tensor. It is necessary to take
this circumstance into account both in development of a theory of propagation
of electromagnetic waves and in analysis of the experimental data. In
particular, near the exchange resonances it is possible to resonantly enhance
such magneto-optic effects as Faraday rotation, the Cotton-Mouton effect, and
the Kerr effect.

It should also be kept in mind that, in addition to the linear processes we
have investigated here, the Hamiltonian (3) also contributes to two-magnon
absorption, to parametric instability, and to other nonlinear phenomena, even
in an approximation linear in the electric field.

The authors wish to express their gratitude to V. G. Bar'yakhtar, V. V.
Eremenko, A. I. Zvyagin, and A. A. Stepanov for discussions of this work.

TABLE I. Magnetic configurations which are irreducible relative to the
crystallographic group $D_{3d}^{6}$ .

\begin{center}%
\begin{tabular}
[c]{|c|c|c|}\hline
{\small Irreducible\hspace{0in}representation} & {\small Irreducible spin
configuration} & {\small Polarization of electric and magnetic fields}\\\hline
$A_{1g}$ & $l_{1z}$ & $-$\\\hline
$A_{2g}$ & $m_{z}$ & $h_{z}$\\\hline
$E_{g}$ & $(m^{_{+}},m_{^{-}})$, $\ \ (l_{1}^{+},-l_{1}^{-})$ & $(h^{+}%
,h^{-})$\\\hline
$A_{1u}$ & $l_{3z}$ & $-$\\\hline
$A_{2u}$ & $l_{2z}$ & $E_{z}$\\\hline
$E_{u}$ & $(l_{2}^{+},l_{2}^{-})$,\ $\ (l_{3}^{+},-l_{3}^{-})$ & $(E^{+}%
,E^{-})$\\\hline
\end{tabular}

\end{center}

TABLE II. Classification of the homogeneous magnetic resonance frequencies in
$\alpha$-Fe$_{2}$O$_{3}$ by symmetry type and components of $\chi_{ij}%
(\omega)$ and $\alpha_{ij}(\omega)$ which have a pole at given AM or EM.

\begin{center}%
\begin{tabular}
[c]{|c|c|c|c|c|}\hline
{\small Ground state} & {\small Type of oscillation} & {\small Variables} &
$\chi_{ij}(\omega)$ & $\alpha_{ij}(\omega)$\\\hline
$A_{1g}$, {\small EA-phase} & AM & $m^{+},l_{1}^{+}$ & $\chi_{xx},\ \chi
_{xy},\ \chi_{yy}$ & $-$\\\hline
& AM & $m^{-},l_{1}^{-}$ & $\chi_{xx},\ \chi_{xy},\ \chi_{yy}$ & $-$\\\hline
& EM & $l_{2}^{+},l_{3}^{+}$ & $-$ & $\alpha_{xx},$ $\alpha_{xy},\ \alpha
_{yy}$\\\hline
& EM & $l_{2}^{-},l_{3}^{-}$ & $-$ & $\alpha_{xx},\ \alpha_{xy},\ \alpha_{yy}%
$\\\hline
$E_{g}$, {\small EP-phase} & AM & $m_{y},m_{z},l_{1x}$ & $\chi_{yy},$
$\chi_{yz},\ \chi_{zz}$ & $-$\\\hline
& AM & $m_{x},l_{1y},l_{1z}$ & $\chi_{xx}$ & $-$\\\hline
& EM & $l_{2x},l_{3y},l_{3z}$ & $-$ & $\alpha_{xx}$\\\hline
& EM & $l_{2y},l_{2z},l_{3x}$ & $-$ & $\alpha_{yy},$ $\alpha_{yz}%
,\ \alpha_{zz}$\\\hline
\end{tabular}

\end{center}

TABLE III. Classification of the homogeneous magnetic resonance frequencies in
Cr$_{2}$O$_{3}$ by symmetry type and components of $\chi_{ij}(\omega)$ and
$\alpha_{ij}(\omega)$ which have a pole at given AM or EM.

\begin{center}%
\begin{tabular}
[c]{|c|c|c|c|c|}\hline
{\small Ground state} & {\small Type of oscillation} & {\small Variables} &
$\chi_{ij}(\omega)$ & $\alpha_{ij}(\omega)$\\\hline
$A_{1g}$, {\small EA-phase} & AM & $m^{+},l_{3}^{+}$ & $\chi_{xx},$ $\chi
_{xy},$ $\chi_{yy}$ & $\alpha_{xx},$ $\alpha_{xy},$ $\alpha_{yy}$\\\hline
& AM & $m^{-},l_{3}^{-}$ & $\chi_{xx},$ $\chi_{xy},$ $\chi_{yy}$ &
$\alpha_{xx},$ $\alpha_{xy},$ $\alpha_{yy}$\\\hline
& EM & $l_{1}^{+},l_{2}^{+}$ & $\chi_{xx},$ $\chi_{xy},$ $\chi_{yy}$ &
$\alpha_{xx},$ $\alpha_{xy},$ $\alpha_{yy}$\\\hline
& EM & $l_{1}^{-},l_{2}^{-}$ & $\chi_{xx},$ $\chi_{xy},$ $\chi_{yy}$ &
$\alpha_{xx},$ $\alpha_{xy},$ $\alpha_{yy}$\\\hline
\end{tabular}

\end{center}

\end{document}